\documentclass[11pt]{amsart}

\usepackage{suffix}
\usepackage{mathtools}
\usepackage{braket}
\usepackage{physics}
\usepackage{tikz}
\usetikzlibrary{arrows}
\usetikzlibrary{calc}

\DeclarePairedDelimiterX\MeijerM[3]{\lparen}{\rparen}%
{\begin{smallmatrix}#1 \\ #2\end{smallmatrix}\delimsize\vert\,#3}

\newcommand\MeijerG[8][]{%
	G^{\,#2,#3}_{#4,#5}\MeijerM[#1]{#6}{#7}{#8}}

\WithSuffix\newcommand\MeijerG*[7]{%
	G^{\,#1,#2}_{#3,#4}\MeijerM*{#5}{#6}{#7}}

\usepackage{amssymb}
\theoremstyle{plain}

\newmuskip\pFqmuskip

\newcommand*\pFq[6][8]{%
	\begingroup 
	\pFqmuskip=#1mu\relax
	\mathcode`\,=\string"8000
	\begingroup\lccode`\~=`\,
	\lowercase{\endgroup\let~}\pFqcomma
	{}_{#2}F_{#3}{\left[\genfrac..{0pt}{}{#4}{#5};#6\right]}%
	\endgroup
}
\newcommand{\pFqcomma}{\mskip\pFqmuskip}

\newtheorem{theorem}{Theorem}[section]
\newtheorem{lemma}[theorem]{Lemma}
\newtheorem{proposition}[theorem]{Proposition}
\newtheorem{corollary}[theorem]{Corollary}
\newtheorem{definition}[theorem]{Definition}
\newtheorem{example}[theorem]{Example}

\newtheorem{remark}[theorem]{Remark}
\newtheorem{conjecture}[theorem]{Conjecture}

\newtheorem{question}[theorem]{Question}
\newcommand \bth[1] { \begin{theorem}\label{t#1} }
	\newcommand \ble[1] { \begin{lemma}\label{l#1} }
			\newcommand {\ethe} { \end{theorem} }
		\newcommand {\ele} { \end{lemma} }
		\newcommand \bpr[1] { \begin{proposition}\label{p#1} }
			\newcommand \bco[1] { \begin{corollary}\label{c#1} }
		\newcommand \bde[1] { \begin{definition}\label{d#1}\rm }
					\newcommand \bex[1] { \begin{example}\label{e#1}\rm }
						\newcommand \bre[1] { \begin{remark}\label{r#1}\rm }
							\newcommand \bcon[1] { \begin{conjecture}\label{con#1}\rm }
								\newcommand \bque[1] { \begin{question}\label{que#1}\rm }
			\newcommand{\beq}{\begin{equation}}
			\newcommand{\eeq}{\end{equation}}
			\newcommand{\beqa}{\begin{eqnarray}}
			\newcommand{\eeqa}{\end{eqnarray}}
			\newcommand{\beaa}{\begin{eqnarray*}}
				\newcommand{\eaa}{\end{eqnarray*}}

		\newcommand {\epr} { \end{proposition} }
						\newcommand {\eco} { \end{corollary} }
					\newcommand {\ede} { \end{definition} }
				\newcommand {\eex} { \end{example} }
			\newcommand {\ere} { \end{remark} }
		\newcommand {\econ} { \end{conjecture} }
	\newcommand {\eque} { \end{question} }

\newcommand \leref[1]{Lemma \ref{l#1}}

\newcommand \deref[1]{Definition \ref{d#1}}

\def \B {{\mathcal B}}

\def \R {{\mathcal R}}
\def \F {{\mathcal F}}

\def \Rset {{\mathbb R}}
\def \Cset {{\mathbb C}}
\def \Zset {{\mathbb Z}}
\def \Nset {{\mathbb N}}
\def \Vset {{\mathbb V}}

\def \diag { {\mathrm{diag}} }

\def \ad { {\mathrm{ad}} }
\def \Re { {\mathrm{Re}} }

\newcommand{\al}{\alpha}
\newcommand{\be}{\beta}

\def \dd {\mathrm{d}}
\def \Rset {{\mathbb R}}         
\def \Cset {{\mathbb C}}

\def \Zset {{\mathbb Z}}

\def \Nset {{\mathbb N}}

\def \B  {{\mathcal{B}}}               

\def \de {\delta}
\def \al {\alpha}
\def \be {\beta}

\def \La {\Lambda}

\def \ga {\gamma}
\def \de {\delta}

\renewcommand \Re { {\mathrm{Re}} }

\newmuskip\pFqmuskip

\usepackage{hyperref}

\begin{document}
\title[d-orthogonal polynomials,  Toda Lattice     and Virasoro symmetries  ]
 {d-orthogonal polynomials,   Toda Lattice   and Virasoro symmetries }
 	
 	\author[E.~Horozov]{Emil Horozov}
 	\address{
 		Department of Mathematics and Informatics \\
 		Sofia University \\
 		5 J. Bourchier Blvd. \\
 		Sofia 1126 \\
 		Bulgaria, and\\	
 		Institute of Mathematics and Informatics, \\ 
 		Bulg. Acad. of Sci., Acad. G. Bonchev Str., Block 8, 1113 Sofia,
 		Bulgaria	}
 	\email{horozov@fmi.uni-sofia.bg}

\date{\today}

\date{}

\begin{abstract} 
	The subject of this paper is  a connection between d-orthogonal polynomials and the  Toda lattice hierarchy. In more details we consider some polynomial systems similar to Hermite polynomials, but satisfying $d+2$-term recurrence relation, $d >1$.  Any such  polynomial system defines a solution of the Toda lattice hierarchy.   However we impose also the condition that the polynomials  are also  eigenfunctions of a differential operator,   i.e. a bispectral problem. This leads to a solution of the Toda lattice hierarchy, enjoying  a number of special properties. In particular the corresponding tau-functions    $\tau_m$   satisfy the Virasoro constraints.   The most spectacular  feature of these tau-functions is that all of them are partition functions of matrix models. Some of them are well  known matrix models - e.g. Kontsevich model, Kontsevich-Penner models,  $r$-spin models, etc. A remarkable  phenomenon is that the solution corresponding to $d=2$ contains  two famous tau functions describing the intersection numbers on moduli spaces of compact Riemann surfaces and  of open Riemann surfaces.

\end{abstract}

\maketitle


\medskip

\section{Introduction}  \label{intro}

\bigskip

In 1975 J. Moser published the  paper \cite{Mo}, in which he made explicit the connection between orthogonal polynomials and the Toda lattice.  Namely, starting with any system of orthogonal polynomials he pointed out that the Toda flow of the coefficients of the 3-term recurrence relation can be expressed in terms of the Stieltjes function (the generating function of the moments of the corresponding measure) of the polynomial system. Later this research was continued both for orthogonal polynomials and for their generalizations (like multiple orthogonal polynomials), see e.g.  \cite{AvM2, AFM, PSZ}.

The present paper also  deals with integrable systems of Toda type and their connections with polynomial systems which are generalizations of orthogonal polynomials.  Before explaining the main results we  give       a brief account of the needed notions.

In 1986 Duistermaat and Gr\"unbaum  \cite{DG} introduced the notion of bispectral operators. The operators $L(x)$ in the variable $x$ and $\La(z)$ in $z$ are called bispectral if  there exists a joint eigen-function $\Psi(x,z)$ such  that 

\[
\begin{split}
L(x)\Psi(x,z) = f(z)\Psi(x,z)\\
\La(z) \Psi(x,z) = \theta(x)\Psi(x,z).
\end{split}
\]
Here   $f(z)$     and    $\theta(x)$     are some nonconstant  functions.    I don't specify the type of the operators, neither the variables as they can be of different types. Some examples  will be more helpful to clarify the situation.

\smallskip

\textbf{1)} $\psi(x,z) = e^{xz}$,  $L= \partial_x$,  $\La = \partial_z$,  $x, z \in \Cset$.

	\[
	L e^{xz} = z e^{xz},  \;\; \La e^{xz} = x e^{xz}.
	\]

\textbf{2)}	Airy operator
\[
\begin{split}
(\partial_x^2 -   x)Ai(x+z)  = zAi(x+z)\\
(\partial_z^2 -   z)Ai(x+z)  = xAi(x+z).
\end{split}
\]

\textbf{3)}	Let T be  the shift operator in $n  \in \Zset$; $T^{\pm 1}f(n)   = f(n\pm 1)$  and $ L= x\partial_x $,   $x \in \Cset$.  Then
\[
L x^n = nx^n, Tx^n = x\cdot x^{n}.
\]

 \textbf{4)} Let $H_n(x), \; n=0, 1,\ldots$ be the Hermite polynomials. Then
	
	\[
	(-\partial_x^2 +x\partial_x) H_n(x) = n H_n(x),\; (T -nT^{-1})H_n(x)  = x H_n(x).
	\]

The bispectral operators have a number of  connections with other fields of research  such as integrable systems (KP hierarchy,
Sato's Grassmannian and the Calogero-Moser systems, orthogonal polynomials, representation theory of $W_{1+ \infty}$ and Virasoro  algebras, 
ideal structure and automorphisms of the first Weyl algebra, etc.  See \cite{BHY1,  BHY2, BHY3, Wil,    BW }

		In fact the examples 2) Airy functions and 4)  Hermite polynomials are obtained via automorphisms of the Weyl algebra    from 1) and respectively from 3). This will be demonstrated   in the next section in more general context.

The  main objects  in the paper  are  the   polynomial systems $P_n(x)$, called $d$-orthogonal polynomials with some extra structure  - they enjoy the generalized Bochner property, see below.     $d$-orthogonal polynomials are
orthogonal with respect to $d$ measures     (see for details \deref{vop})   rather than one. Then using these polynomials we are going to find special solutions of the Toda lattice hierarchy which lead  to important matrix models   as explained in Section 8.

As in the case of orthogonal polynomials this property has purely algebraic expression.     Due to   P. Maroni  \cite{Ma},  we can use as a definition of   
$d$-orthogonality   the following one: the polynomials satisfy a $d+2$-term recursion relation of the form

\[
x\cdot P_n(x) =  P_{n+1}(x) + \sum_{j=0}^{d}  \ga_j(n) P_{n-j}(x).
\]
with constants $\gamma_j(n)$, independent of $x$, $\ga_d(n) \neq 0$  for   $n\geq d$.

\bigskip

 \textbf{Generalized Bochner problem (GBP).}    
Find systems of polynomials $P_n(x), \; n=0, 1, \ldots$ that are eigenfunctions of a differential  operator $L$ of order $m$ with eigenvalues $\lambda(n)$ depending on the discrete variable $n$ (the index):

\[
LP_n(x) = \lambda(n)P_n(x)  
\]
and which at the same time are eigenfunctions of a difference operator, i.e. that satisfy a finite-term (of fixed length $d+2$), recursion relation of the form

\[
x\cdot P_n(x) = P_{n+1}(x) + \sum_{j=0}^{d}\gamma_j(n)P_{n-j}(x).
\]
In the case of classical orthogonal polynomials Bochner's theorem shows that   all their properties follow from the   fact that they are eigenfunctions of a differential operator.  The  above generalization produces  $d$-orthogonal polynomials analogs of the classical orthogonal polynomials, which have very similar properties.

We found a large class of d-orthogonal polynomials with this property  (conjecturally all).   See Section 2 or \cite{Ho1} for details.


\bigskip

The construction goes as follows. We start with Weyl algebra $W_1$, i.e.     the algebra of  differential operators in one variable $x$ with polynomial coefficients.   It has a natural representation in the space of all polynomials $\Cset[x]$.

  Consider  the trivial polynomial system $\{x^n\}, \; n=0, 1,2,\ldots$.   The polynomials $x^n$ are eigenfunctions of  the differential operator $H =x\partial_x$ with eigenvalues $n$. They also satisfy the trivial recurrence relation

\[
x\cdot x^n = x^{n+1}  = T x^n,
\]
where $T$ is the shift operator $Tf(n) =f(n+1)$.  Also notice the identity $\partial_x x^n =n x^{n-1}$.  We see  that the operators $H$ and $T$ are bispectral.  Now we will explain how to construct non-trivial solutions of the bispectral problem.

It is known (and easy to see), cf. \cite{Dix}, that  given a polynomial $q(\partial_x)$ without constant term, one can construct an automorphism of $W_1$ by the formula

\[ 
\sigma A :=e^{\ad_{q(\partial_x)}} A = \sum_{j=0}^{\infty} \frac{\ad_{q(\partial_x)}^jA}{j!},  \;\; A\in W_1,
\]
where   $\ad_B A = [B,A]$     and    $\ad_B^{j} A = [B,    \ad^{j-1}_B A]$ .
Then following \cite{BHY1} we can use $\sigma$ to obtain  a new pair of bispectral operators with non-trivial polynomial system $P_n(x)$, being the common eigenfuction. Namely we put

\[
P_n(x)   = \sum_{j=0}^{\infty} \frac{q(\partial_x)^j   x^n}{j!}.
\]
It is obvious that the above sum is finite and $P_n(x)$ is a polynomial of degree $n$. Also introduce the operators
\[
L=\sigma H, \;\; D = \sigma^{-1}x.
\]
For example if $q(\partial_x)  = - \partial_x^2/2$ we obtain that 

\[
L = -\partial_x^2 +x\partial_x  \;\; \sigma \partial_x   = \partial_x\;\; \text{and} \;\; \sigma^{-1} x = x + \partial_x^2.
\]
Then 

\[\begin{split}
LP_n(x) & = nP_n(x)\\
xP_n(x)& = [T +n(n-1)T^{-1}] P_n(x)
\end{split}
\]
For the last identity, see \cite{Ho1} or the next section for details.  Thus  we obtain the classical Hermite polynomials.  The case of general  polynomial $q$ is treated in the same manner.  In this paper we consider the special case of $q(\partial_x)  = \partial_x^{d+1}/(d+1)$,  $d\in \Nset, \; d\geq 1$. Notice that we obtain simultaneously  all ingredients - the polynomial system, the differential operator and the recurrence relation. 

Our next step is to construct the measures or which is equivalent - their weights.  This is done in Section 3. It turns out that they play the same role as the polynomials in the construction of solutions of Toda hierarchy. This can be seen from their integral representations, which are of the same class as the polynomials:

\[
\nu(n, x)  = \int_{C}z^{-n-1} \exp(\frac{-z^{d+1}}{d+1}  +xz) \dd z
\]
with an appropriate contour $C$. The polynomials correspond to closed contour around $z=0$ and $n= 0, 1,\ldots$. For the weights we have to choose a contour, going to infinity in both directions, on which the integrand tends to zero and $n=-1, -2, \ldots$. However the above integrals make sense for all values of $n\in \Zset$ when we chose the contour as for the weights. Moreover the set of functions $\nu(n,x), \; n\in \Zset$ satisfy both the recurrence relations and the differential equation, i.e. they give another solution of the bispectral problem, this time the range of the variable $n$ is $\Zset$. Let us formulate the corresponding result, proven in Section 4.

\ble{diff}
(i)  The functions $\nu(n,  x) $ satisfy the equation

\beq  \label{diffeq}
- \partial_x^{d+1}\nu(n, x) +x\partial_x \nu(n, x) = n\nu(n, x).
\eeq

(ii) They  satisfy the recurrence relation

\beq \label{recur}
x\nu(n, x) =   (n+1)      \nu (n +1,x)+ \nu(n-d, x).
\eeq

(iii)  The lowering operator  $\partial_x$ acts on the solutions $\nu(n, x)$ as

\beq \label{low}
\partial_x \nu(n, x)  =  \nu(n-1, x).
\eeq
\ele

In the theory of classical orthogonal polynomials   functions with such properties are called functions of the second kind  \cite{Seg}. Each contour $C$ for which the integral is convergent defines such type of functions.   

This gives a bi-infinite system of functions, from which we can  construct a solution of the the Toda hierarchy.

We know (after Sato), see \cite{Sa,  JM} that both Toda and KP hierarchy of equations  can be written as evolution equations in a certain infinite-dimensional Grassmannian - Sato Grassmannian. Each point of the Sato's Grassmannian  corresponds to a plane  $W$, which consists of functions in  an infinite-dimensional linear space.   We will be interested in planes possessing  certain symmetries.

Let us recall the  idea of these symmetries on a simple example.    In the studies of generalized Kontsevich models it turned out that a very important ingredient is the notion of   Kac-Schwarz   \cite{KS} operators   $A_j$. Roughly speaking they act on a solution  of Airy equations, spanning the entire plane  $W$  and leaving it invariant:  $A_jW \subset W$.

The above constructed functions $\nu(n, x)$ span a flag    $F = \ldots W_n \subset W_{n+1} \subset \ldots$ of planes $W_n  = (\nu(n, x),    \nu(n-1, x),  \ldots)$. Each of these planes gives rise to a tau function  $\tau(n, t)$,   where $t = (t_1, t_2, \ldots)$ via the boson-fermion correspondence.


In the simplest case of $d=2$ the  differential operator is 

\[
L = - \partial_x^3 + x\partial_x.
\]
It is easy to check that  the corresponding function $\nu(-1, x)$ is   the Airy function $Ai(x)$.   It is the first weight function, defining the measure for  the $d$-orthogonal 
polynomials    $P_n(x)$, $d=2$, which are eigenfunctions of the differential operator. 
The rest of the weights are given by the  derivatives of $Ai(x)$.  

The next result in the paper  shows that all the tau-functions satisfy the Virasoro constraints. Virasoro algebra is an algebra spanned by operators $L_j , j\in \Zset$ and the central element $c$ and satisfying the commutator relations

\[
[L_k, L_m]  = (k -m) L_{k+m} \frac{k^3 -k}{12} \de_{k,-m}c.
\]
In the bosonic representation the operators $L_j$ are differential operators of order one or two in infinite number of variables $t_1, t_2, \ldots$.
The Virasoro constraints are equations of the type

\[
L_j\tau(n,t) = 0, \;\text{for} \; j=-1, 1,2, \ldots
\]
while the function $\tau(n, t)$ is an eigenvector of the operator $L_0$.   For the case of    $d=3$ and $n=-1$ the result belongs to   Witten and  Kontsevich \cite{Wi, Kon} (although in different terminology and notations)   For $d=3$ and $n=0$ the above result belongs to Alexandrov \cite{Al2}.

The corresponding tau-function $\tau(-1, t)$ has a deep and beautiful algebro-geometric meaning as  conjectured  by  Witten  \cite{Wi}  and proved by  Kontsevich  \cite{Kon}.  Namely it gives the intersection theory of  the moduli spaces of compact Riemann surfaces.

The function $\tau(0, t)$ also has algebro-geometric meaning as proved by Alexandrov, Buryak, Tessler \cite{Al2,  BT}.  It describes the intersection theory of moduli spaces of  open Riemann surfaces.   See the cited papers and the  references therein.

The above tau-functions have appeared in the cited references as solutions of other integrable hierarchies - KP or extended KP. Here  they  appear as \textit{one solution }  of the Toda lattice $\tau(n, t)$. 

For the other values of $d$ at least one of the tau-functions also has algebro-geometric meaning.  For $d>2$ the function $\tau(-1, t)$  is connected with the so called $r$-spin structures, introduced by Witten   \cite{Wi}, see also  \cite{FSZ}.

Let us point that the connection of  the bispectral problem with the   solutions of integrable systems exhibiting Kac-Scwartz symmetry has been discussed in many different places. The first documented one that I know is in \cite{AvM1}. Here we present other instances of such connections.
\subsection*{Acknowledgements}

This research has been partially supported by the Grant No DN 02-5 of the Bulgarian Fund "Scientific research".

\medskip


   


 \section{Preliminaries} \label{pre}

 \subsection{$d$-orthogonal polynomials}

 \bde{vop}
  Let $\{P_n(x),\;  n= 0, 1, \ldots\}$ be a family of monic polynomials such that $\deg P_n = n$. The polynomials are d-orthogonal iff   there exist $d$  functionals $\mathcal{L}_j, j =0,\ldots,d-1$    on the space of all polynomials  $\Cset[x]$   such that
  
  \[
  \begin{cases}
  \mathcal{L}_j (P_nP_m)    = 0, m > nd+ j, n \geq 0,  \\
  \mathcal{L}_j(P_nP_{nd+ j})    \neq 0, n   \geq 0,    
  \end{cases}
  \]
  for each $j \in N_{d+1} := \{0, \ldots, d-1 \}$.    When $d = 1$ this is the ordinary notion of orthogonal polynomials.  
 
 \ede
 
  Notice that   $  \mathcal{L}_j(P_{j})    \neq 0, \; j= 0, \ldots, d-1$. 
 The notion of $d$-orthogonal polynomials was introduced by J. Van Iseghem,   \cite{VIs}.

    Let $v(x)$ be the   weight (function) on a subset $U \subset \Rset$, which defines the functional 
   \[
   \mathcal{L}(P) = \int_U v(x)P(x) \dd x.
   \]
   Instead of $ \mathcal{L}(P)$    we are going to use the notation

 \[
 \left\langle v(x),    P(x)    \right\rangle,
 \]
 which is very convenient for algebraic manipulations.

   One can show easily that for each $n$ there exist polynomials 
 $A_0^{(n)}(x), \ldots A_{d-1}^{(n)}$ such that the weights  $v_n = \sum_{i=0}^{d-1} v_j A_j^{(n)}$ have the property

 \[
 \left\langle v_j(x),    P_m(x)    \right\rangle = \de_{j,m}c_j \; c_j \neq 0.
 \]

The orthogonality connected with    $d$ functionals rather than with only one, gives the  name of  \textit{d-orthogonal polynomials}.

 Norming suitably the weights $v_j$ we see that they define    the bilinear forms
 with the properties

 \[
 \left\langle v_j(x),    P_m(x)    \right\rangle = \de_{j,m}.
 \]

 A very important theorem of P. Maroni \cite{Ma} is the following one.
 
 \bth{VIs}
 A polynomial system $\{P_n(x)\},\;  n\geq 0$, is d-orthogonal if and only if  the polynomials satisfy a $d+2$-term recurrence relation of the form
 
 \beq  \label{d-ort}
 xP_n(x) = P_{n+1} + \sum_{j=0}^{d}\gamma_j (n)P_{n-j}(x)
 \eeq
 with constants $\gamma_j(n)$, independent of $x$, $\ga_d(n) \neq 0, n \geq d$. 
 \ethe

 \subsection{Bochner's property}

 In this subsection we briefly recall some   notions and   results from \cite{Ho1, Ho2} about d-orthogonal  polynomials, which are eigenfunctions of a differential operator. We have called this property       "Bochner's property"   in  analogy with the classical theorem by Bochner    \cite{Bo}.

 The differential operators with polynomial coefficients form  the   Weyl algebra $W_1$.  	
 It is  spanned   by the two generators $x, \partial_{x}$.

 \bigskip

 For any polynomial $q(\partial_x)    =  \sum_{j=1}^{d+1} a_j \partial_x^j \in \Cset[\partial_x]$ (notice that the coefficients $a_j$ are constants)  we define the automorphism  of $W_1$
 
 \[
 \sigma_q(A) = e^{\ad_{q(\partial_x)}}(A)  =  e^q A  e^{-q}, \;\; A \in W_1.
 \]
 Here $\ad_A(B) = [A, B]   =AB- BA$,   $A,B \in W_1$.


 For any polynomial $q(\partial_x)    =  \sum_{j=1}^{d+1} a_j \partial_x^j$ without a constant term\footnote{The constant term contributes only to multiplication of the polynomials by a constant. On the other hand without it the formulas and the arguments are simpler.} we defined the automorphism $\sigma_q = e^{q(\partial_x)}$ of $W_1$.

 Then we found   the images of the generators $x,  \partial_x$   of $W_1$ under the automorphism $\sigma_q$:
 
 \ble{aut}
 
 \begin{equation}
 \begin{cases}
 \sigma_q(\partial_x) = \partial_x\\
 \sigma_q(x) = x + q'(\partial_x).
 \end{cases}  \label{aut}
 \end{equation}

 \ele
 
 \qed

 Let us introduce an auxiliary  algebra $\R_2$   defined over  $\F$    with generators    $T, T^{-1}, \hat{n}$  subject to the      relations 
 
 \[
 T\cdot T^{-1} = T^{-1}\cdot T=1, \; [T, \hat{n}] = T, \; [T^{-1}, \hat{n}] = -T^{-1}.
 \]
 One easily sees that   $[T, \hat{n} T^{-1}] = 1$. Hence the operators $T,   \hat{n} T^{-1}$ define a realization of the Weyl algebra. We can   introduce another algebra $\B_2$ as follows.

 First we define an anti-homomorphism $b$, i.e. a map $b\colon W_1 \to \R_2$ satisfying 
 
  \[
  b(m_1 \cdot m_2) = b(m_2)\cdot b(m_1),
  \]
   for each $m_1,\; m_2 \in W_1$  by:
 
 \[
 \begin{cases}
 b(x)= T\\
 b(\partial_x) = \hat{n}T^{-1}\\
 \end{cases}
 \]
 The algebra   $\B_2$ will be the image $  b(W_1)$   of $W_1$. Then  $b\colon W_1 \to \B_2$ is an anti-isomorphism and in particular $b^{-1} \colon      \B_2 \to W_1$ is well defined.
 
  We start with the  polynomial system  $\psi(x, n) = x^n$.  Then the corresponding operators $\partial_x, \hat{x}$ act on $\Cset[x]$.       In the same way we represent the algebra $\B_2$ in $\Cset[x]$   by realizing $T$ and $T^{-1}$   as the shift operators  acting on functions $f(n)$ by $T^{\pm}f(n)     = f(n\pm 1)$.   Finally $\hat{n}$ denotes the operator of  multiplication by $n$.
 With this notation we have

 \ble{bisp}
 
 \[
 \begin{cases}
 \partial_x\psi(x,n) = \hat{n}  T^{-1} \psi(x,n)\\
 x \psi(x,n) =  T \psi(x,n)\\
 x\partial_x \psi(x,n)  =  n\psi(x,n).
 \end{cases}
 \]
 \ele

 \qed

 Using $q(\partial_x)$ we   define another polynomial system $P_n(x)$ by

 \[
 P_n(x) = e^{q(\partial_x)}\psi(x,n)= \sum_{j=0}^{\infty}\frac{q(\partial_x)^j\psi(x,n)}{j!}.
 \]
 Notice that the above series is in fact finite as the operator  $q(\partial_x)$ reduces the degree of any polynomial. Let us denote the operator $\sigma_q(x\partial_x)$ by $L$. We proved in \cite{Ho1}     among other statements the following

 \bth{any}
 The polynomials $P_n(x)$ have the following properties:

 (i) They are eigenfunctions of the differential operator 
 
 \begin{equation}      \label{new-diff3}
 L :=  q^{'}(\partial_x)\partial    + x\partial  
 \end{equation}
 with eigenvalues $\lambda(n)  =n$.
 
 (ii) They have the lowering operator   $\partial_x$, i.e.

 \begin{equation}     \label{new-low}
 \partial_x P_n(x)  =   nP_{n-1}. 
 \end{equation}
 
 (iii) They satisfy  the  recurrence    relation  
 
 \begin{equation}     \label{new-rec3}
 xP_n(x)  =P_{n+1} +\sum_{j=0}^{d}ja_j   n(n-1)\ldots (n-j+1)P_{n-j}. 
 \end{equation}
 
 \ethe


 \proof

 We   repeat the  proof from \cite{Ho1} as here it  is simpler.

 \[  (i)  \;\; L P_n(x)  = ( e^{q(\partial)}\cdot x\partial  \cdot e^{- q(\partial)})\cdot (e^{q(\partial)} x^n) =   e^{q(\partial)}x\partial  x^n = n P_n(x)
 \]
 
 (ii)  		is obvious.

 (iii)				
 This  requires  more computations but it is similar to (i). One has to compute $\sigma^{-1} x = e^{-q(\partial)}\cdot x \cdot e^{ q(\partial)}$ and then use the obvious $x\cdot x^n = x^{n+1}$.  Then the expression $\sigma^{-1} x\cdot x^n$   has to be  written only in terms of $T$ and $nT^{-1}$, using that $x\cdot x^n   = x^{n+1}$ and $\partial x^n = n x^{n-1}$. In details the computation is:		
 \[
 \begin{split}
 &	   x \cdot P_n(x)       =  x\cdot  e^{q(\partial)} x^n 
 = e^{q(\partial)}  e^{- q(\partial)}  \cdot x\cdot  e^{q(\partial)} x^n\\
 &= e^{q(\partial)} (\sigma^{-1} x) \cdot x^n =
 e^{q(\partial)} \big[ x + \sum_{j=0}^{d}\ga_j \partial^j   \big] \cdot x^n\\
 &	   	       e^{q(\partial)} \cdot\big[ x^{n+1} + \sum_{j=0}^{d}\ga_j n(n-1)\ldots (n-j+1)x^{n-j}   \big]. 
 \end{split}
 \]
 
 \qed

 \begin{example}

 	$q(\partial_x)   = - \partial_x^{d+1}/(d+1) $ corresponds to Gould-Hopper polynomials \cite{GH}. When $d=1$   these are the Hermite polynomials.  The operator $L$ reads
 	
 	\[
 	L   =    	 x\partial_x   -\partial_x^{d +1}. 
 	\]
 	For the recurrence  relation we need to compute $\sigma^{-1} x$. Similarly to the above we obtain that 
 	
 	\[
 	\sigma^{-1} x = x+ n(n-1)\ldots (n-d)\partial^{n-d},
 	\]
 	which gives
 	\[
 	xP_n  =  P_{n+1}  + n(n-1)\ldots(n-d)P_{n-d}.
 	\]			
 	
 \end{example}
 \vspace{0.3cm}

\section{Weights}

\subsection{Pearson  equations}

In what follows  with an abuse of language we will not distinguish between a functional  and its weight  $v (x) $.  Let    $\{P_j(x)\}, \; j=0, 1, \ldots$  be a system of polynomials $\deg P_j(x)   =j$.      We introduce the dual system of functionals    $v_j(x) $      on $\Cset[x]$, such that 

\[
\left\langle v_j,  P_n(x) \right\rangle  =\de_{jn}, \;\;      j,n =0, 1 \ldots.
\]
For each functional $v$ we can define new functionals $xv$  and $\partial_x v$ by

\[
\left\langle xv,  P_n(x) \right\rangle  =   \left\langle v,  xP_n(x) \right\rangle,\;
  \left\langle \partial_x v, P_n(x) \right\rangle   =  \left\langle  v, -\partial_x P_n(x) \right\rangle.
\]

Below we will determine explicitly  the dual system  $v_j(x)$  for the polynomials   $\{P_j(x)\}$, defined by $q(\partial_x)$.

\ble{Pear}
(i)    The functional $v_0(x)$ is a solution of the  differential equation

\beq \label{meas}
\big(  q'(-\partial_x )+  x\big)v_0 = 0.
\eeq

(ii)  The functionals $v_j(x)$ are given by

\[
v_j (x)= \frac{(-1)^j}{j!}\partial^j v_0(x).
\]

\ele

\proof 
From the properties  \eqref{low}   of the polynomials $P_n(x)$ we see that

\[
 \frac{1}{n+1}\partial_x P_{n+1}(x)   =    P_n(x).
\]

(i)   We start with the functional $v_0$. It satisfies  

\[
\left\langle v_0,  P_n(x) \right\rangle  =\de_{0n}.
\]
Let us use the differential operator $L$.    We have $LP_n(x)  = nP_n(x)$.  We will compute the action of $v_0$ on the system $P_n(x)$.

Denote by $L^*$ the formal adjoint  to $L$, i.e. $L^*$ satisfies the identity

\[
\left\langle f, L Q(x)    \right\rangle    =   \left\langle L^* f,  Q(x)    \right\rangle 
\]
for any     polynomials     $Q(x)$ and $f(x)$. Using that    $LP_n(x) =  nP_n(x)$ we obtain

\[
\left\langle v_0, L P_n(x)    \right\rangle   =\left\langle v_0, nP_n(x)    \right\rangle = 0, \;\; \text{for all} \;\; n\geq 0.
\]
Hence 

\[
0 = 	\left\langle L^*v_0, P_n(x)    \right\rangle = 0, \;\; \text{for  all} \;\; n\geq 0,
\]
where   the operator $L^*$ explicitly reads

\[
L^* = -\partial_x q'(-\partial_x ) - \partial_x x.
\]
We see that the functional $ L^*v_0 \equiv 0$.  Integrating once   the function   $      \partial_x [q'(-\partial_x ) + x]v_0(x) =0 $         we obtain that

\[
\big(  q'(-\partial_x ) + x\big)v_0 = c, \; \; c\in \Cset.
\]
We will compute the constant by applying  the functional $\big(  q'(-\partial_x +  x\big)v_0 $ to the system $P_n(x)$,    $n=0, 1, \ldots$. We have

\[
\begin{split}
&\left\langle   c, P_n(x)    \right\rangle  =\\
& = \left\langle \big(  q'(-\partial_x )+ x\big)v_0, P_n(x)    \right\rangle \\
&=     \left\langle \big(  q'(-\partial_x )+  x\big)v_0,   \frac{1}{n+1} \partial_x P_{n+1}(x)    \right\rangle \\
&=   \left\langle v_0,   \frac{1}{n+1}  L P_{n+1}(x)    \right\rangle \\
&=    \left\langle v_0,   \frac{n+1}{n+1}   P_{n+1}(x)    \right\rangle =    \left\langle v_0,     P_{n+1}(x)    \right\rangle =0.
\end{split}
\]
Hence $c=0$ and we see that $v_0(x)$ satisfies the equation \eqref{meas}.

Next  consider
 \[
1 =\de_{00}=  \left\langle v_0,     P_{0}(x)    \right\rangle =   \left\langle v_0,    \partial_x P_{ 1}(x)   \right\rangle =  \left\langle -\partial_x v_0,     P_{1}(x)    \right\rangle,
 \]
 i.e. 
 
 \[
 \left\langle -\partial_x v_0,     P_1(x)    \right\rangle =   1.
 \]
 Using that 
 
 \[
 \de_{0n}=  \left\langle v_0,     P_{n}(x)    \right\rangle =      \left\langle  -\partial_x v_0,     P_{n+1}/(n+1)(x)    \right\rangle = 0
 \]
 for $n>0$ we obtain that  $v_1 = -\partial_xv_0$. We continue by induction to see

\[
v_j (x)= \frac{(-1)^j}{j!}\partial^j v_0(x).
\]

\qed

The functions       $v_j(x)$  are similar to  Airy function and also  have explicit representations in terms of definite integrals. We shall derive them  in the next section.

Using the   bi-orthogonality property and the recurrence relation for the polynomials it is easy to show that the functions    $v_j$ also satisfy recurrence relation of the form:

\[
xv_m(x) =    v_{m-1}(x) +\sum_{j=0}^{d-1} b_{j}(m)v_{m+j}(x),
\]
see e.g. \cite{DK, SVI} for details and more general results. 
 

 \section{Integral representations}

\subsection{Integral representations for the weights}
We will find integral representations of the weights of the polynomials corresponding to     $q(\partial_x)    = -\partial_x^{d+1}/(d+1)  $. As we know,  it is enough to find one for  $v_0(x)$.

\ble{int-Ai}
The weight $v_0(x)$   is given by the formula

\[
v_0(x)    =   \int_C \exp (\frac{-z^{d+1}}{d+1} + xz ) \dd z.
\]
with an appropriate contour $C$,   see Fig. 1.
\ele

\proof

Denote by   $\hat{v}_0(z)$  the   Laplace  transform of $v_0(x)$, i.e.

\[
\hat{v}_0(z) = \int_C e^{-zx} v_0(x)\dd x.
\]
Then  performing the  Laplace  transform on the equation \eqref{meas},  where $q(z) = \frac{-z^{d+1}}{d+1}$,
we obtain 
\[
[ - z^{d}-     \partial_z]\hat{v}_0(z) = 0.
\]
Solving the differential equation we obtain (up to a multiplicative constant)

\[
\hat{v}_0(z) =  \exp (\frac{-z^{d+1}}{d+1} ).
\]
Finally   the inverse   Laplace transform applied to $\hat{v}_0(z)$  gives 

\[
v_0(x)  =  \int_C \exp (\frac{-z^{d+1}}{d+1} + xz ) \dd z,
\]
where the contour $C$ is chosen  so that the integral to be convergent.   All such contours      give solutions.   For example on fig. 1 we have taken   the union of two rays  $R_1 \cup R_{-1}$, where

\[
R_{k}= R \exp( \pm 2\pi i      /(d+1)), \; k=  \pm 1   \;\;  \text{and}\;\; R \in [0, \infty).
\]
We see that on the rays  $R_k$   the function     $\Re\big[-\frac{z^{d+1}}{d+1} +xz\big] = -R^{d+1}(1/(d+1) + \mathcal{O}(R^{-1}))$, as a function of $z$.  Hence the integral is convergent.

The rest of the weights are given by

\beq \label{formula GH mes}
v_j(x)    =  \frac{(-1)^j}{j!}  \int_C z^j \exp (\frac{-z^{d+1}}{d+1}  + xz) \dd z,
\eeq
which follows from \leref{Pear} (ii).

\qed

\begin{figure}
	\begin{tikzpicture}

	\draw[thick,  ->] (0, 0) -- (6,2);
	\draw[thick, ->]  (0, 0) --(6, -2);
	 
	\draw[->] (0,0) -- (5.5,0) node[anchor=north west]{x axis};
	\draw[->] (0,0) -- (0,2.5) node[anchor=south east]{y axis};
	\end{tikzpicture}
	\caption{The integration contour $C$.}
\end{figure}
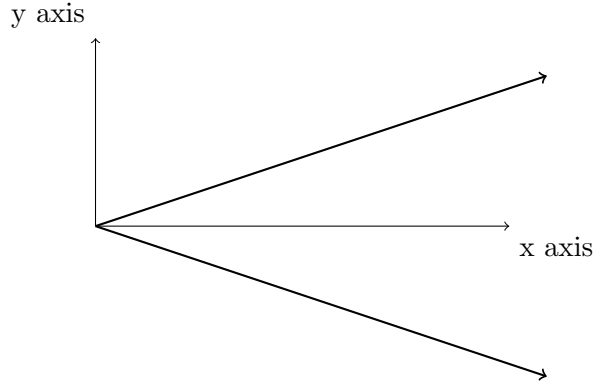


\subsection{Integral representations for the  polynomials}

 Consider the following scalar integrals with suitable contours $C$

 \beq \label{sec-kind}
 w_{n}(x)  =   \int_C z^{-n-1} \exp\big(\frac{-z^{d+1}}{d+1} + xz        \big)    \dd z,  \;\;  n \in \Zset_+.
 \eeq

 \ble{equ-N}
 The function   $w_{n}(x) $   satisfies the equation
 
 \beq  \label{int-pol}
 \big( -\partial_x^{d+1} +x\partial_x\big) w_{n}(x)  = n w_{n}(x).
 \eeq
 
 \ele 
 
 \proof 
 It is an easy computation using integration by parts.   On one hand we have that

 \[
- \partial_x^{d+1}w_n(x) = \int_C z^{-n-1}(-z^{d+1})\exp(\frac{-z^{d+1}}{d+1} +zx)\dd z. 
 \]
 On the other hand integrate by parts the expression
 \[
 \begin{split}
 x\partial_x w_{n}(x)& =  \int_C z^{-n}\exp(\frac{-z^{d+1}}{d+1} +zx) \dd (x  z ).  \\
 \end{split}
 \]
 This gives

 \[
 \begin{split}
 &= \int_C z^{-n}\exp(\frac{-z^{d+1}}{d+1})\dd \exp(zx) \\
 &=- \int_C [(-n)z^{-n-1} -z^{d}z^{-n}]     \exp(\frac{-z^{d+1}}{d+1}+ zx)\dd z\\
 &= \int_C [nz^{-n-1}+ z^{d+1}z^{-n-1}]     \exp(\frac{-z^{d+1}}{d+1}+ zx)\dd z\\
 \end{split}.
 \]
 Hence 
 
 \[
 (-\partial_x^{d+1}  +x\partial_x)w_{n}(x)   = nw_{n}(x).
 \]

 \qed
 
 In fact the different solutions of this differential equation are given by appropriately choosing the contour.   This is done  in a similar way to the case of the weights.  The only thing  we need here is to avoid the point $z=0$  as the integrals are not convergent in $z=0$     when  $n+1 >0$. It can be achieved by
 appropriately  deforming  the contour $C$ to avoid the singularity.   
 
  If we want to obtain representation for the polynomials we need to go around the point $z = 0$.        
 
 \bco{int-pol}
 The monic polynomials $\{P_n(x)\}$   have the following integral representation
 
 \[
 P_n(x)   =   \frac{ n! }{2\pi i}  \int_{|z| =\varepsilon}z^{-n-1}\exp(\frac{-z^{d+1}}{d+1} +zx)\dd z. 
 \]
 
 \eco
 
 \proof
 The polynomial $P_n(x)$ is  the residue of the integral. By expanding the exponent we see that the highest degree $x^n$ of the coefficient at $z^{-1}$ is given by      $x^n/n!$ which gives the needed result.

 \qed
 
 \subsection{Functions of the second kind}
 
The rest of the solutions of the  differential equations for the  classical orthogonal polynomials share  a lot of properties with the polynomial solutions. 
See, e.g. \cite{Is},   sect. 3.6 and \cite{Seg}, 4.23. Here we shall give analogs of these results for the cases of interest.  We are going to prove slightly more general results, taking any  number $s \in \Cset$   instead of non-negative integer $n$ . Let us introduce the functions

\beq \label{eps}
\nu(s,  x)   =  \int_{C}z^{-s-1}\exp(-z^{d+1}/(d+1)+ zx)\dd z,     
\eeq
where  $C$ is a  contour   along which  $\Re\big[(-z)^{d+1}/(d+1)-xz\big] \to -\infty$ and which does not pass through the point $z = 0$.

\ble{recur}
(i)  The functions $\nu(s,  x) $ satisfy the equation

\beq  \label{diffeq-n}
- \partial_x^{d+1}\nu(s, x) +x\partial_x \nu(s, x) = s\nu(s, x).
\eeq

(ii) They  satisfy the recurrence relation

\beq \label{recur-n}
x\nu(s, x) =   (s+1)      \nu (s +1,x)+ \nu(s-d, x).
\eeq

(iii)  The lowering operator  $\partial_x$ acts on the solutions $\nu(s, x)$ as

\beq \label{low-n}
\partial_x \nu(s, x)  =  \nu(s-1, x).
\eeq
\ele

\proof
(i)  The proof repeats the polynomial case.

(ii)  Integrating by parts the expression $x\nu(s, x)$   we obtain

\[
\begin{split}
&  \int_{C}z^{-s-1}\exp(\frac{-z^{d+1}}{d+1})\dd \exp(+xz) =  \\
& =   -\int_{C} [(-s-1)z^{-s-2}   - z^{-s-1  +d}  ] \exp(-\frac{z^{d+1}}{d+1} +xz)\dd  z \\
& =  \int_{C} (s+1) z^{-s-2} \exp(-\frac{z^{d+1}}{d+1} +xz)\dd z \\
& + \int_{C}        z^{-s +d-1}    \exp(-\frac{z^{d+1}}{d+1} +xz)\dd z =\\
& =      (s+1)      \nu(s, x)   +  \nu(s- d, x).
\end{split}
\]

(iii) Differentiating the integral gives

\[ 
\partial_x  \nu(s, x)   =  \int_{C}z^{-s}\exp(-\frac{z^{d+1}}{d+1}+zx)\dd z =  \nu(s-1, x).  \\
\]

\qed

\bre{sec-kind}     The functions $\nu(n, x) $,   where $ n$ is a negative integer correspond to the weights, while $n\geq 0$  correspond to the polynomials (when we take a closed contour around $0$) or to the functions of the second kind. In this way formula \eqref{eps}  gives  integral representations of the same kind both for  the weights and the functions of the second kind (the polynomials).

We can  rescale $\nu(s,x)$   introducing $u(s,x)$   by $\nu(s,x)   = a(s)u(s, x) $ so that the recurrence    \eqref{recur-n}  takes the original form

\[
xu(s,x) = u(s+1,x)  +   b(s- d)u(s- d,x).
\]
In this way we will obtain the recurrence for the monic polynomials $P_n(x)$.
  However we will work primarily  with   \eqref{recur}.
 \ere

 \section{Preliminaries on infinite dimensional Lie algebras}

 \subsection{The Lie algebra $\mathfrak{gl}_{\infty}$    and its representations}

 By $\mathfrak{ gl}_{\infty}$  we denote the algebra of doubly infinite matrices with only finite number of nonzero elements.
We will need also the algebra $\bar{a}_{\infty}$ defined by the condition that the nonzero diagonals are finitely many:

\[
\bar{a}_{\infty} = \{\big( a_{ij}\big)|  a_{ij}  =0 \;\; \text{for} \;\;   |i-j| >>0    \}.
\]

 The fermionic Fock space is defined as follows.    Introduce a vector space $\Vset$   with a fixed basis  $v_j,  \; j \in \Zset$:
 
 \[
 \Vset  = \mathop{ \oplus}_{j \in \Zset} \Cset v_j.
 \]
 The algebra  $\mathfrak{gl}_{\infty}$ acts on $\Vset$ in  the standard way.

 Consider the vectors of the form 
 
 \[
 v_{i_0}\wedge v_{i_{-1} }\wedge v_{i_{-2} }\wedge \ldots
 \]
 such that $i_0 >i_{-1}  > i_{-2} \ldots$ and $ i_{-k} = -k$ for $k$ big enough. We consider the space $F^{(0)}$   spanned by the above vectors with coefficients in $\Cset$.  The element 
 
  \[
 \ket{0} = v_0\wedge v_{-1} \wedge v_{-2} \wedge \ldots
 \]
 is called the   $0$-vacuum.
 
 Exactly in the same way we define the $m$-th vacuum by
 
 \[
 \ket{m} = v_m\wedge v_{m-1} \wedge v_{m-2} \wedge \ldots
 \]
 and the corresponding space $F^{(m)}$. The direct sum 
 
 \[
 F = \mathop{\oplus}_{m \in \Zset} F^{(m)}.
 \]
 will be called fermionic Fock space.

 Next we  define an action of  $\mathfrak{ gl}_{\infty}$  on $F$ as follows. Let $a \in  \mathfrak{ gl}_{\infty}$.  Take a vector $v_{i_0} \wedge v_{i_1} \wedge \ldots$  and  define the map       $r(a) \colon F^{(m)} \to F^{(m+1)}$  as 
 
 \[
 r(a) (v_{i_0} \wedge v_{i_1} \wedge \ldots)  = av_{i_0} \wedge v_{i_1} \wedge \ldots + v_{i_0} \wedge a v_{i_1} \wedge \ldots + \ldots
 \]
 and by linearity.

 We also will need the group-like operator $R(\exp(a))\colon F^{m} \to F^{(m)}$, where $a\in \mathfrak{ gl}_{\infty}$. It is defined by the action

 \[
 R(\exp(a)(v_{i_0} )\wedge v_{i_1} \wedge \ldots)  = \exp(a)v_{i_0} \wedge\exp(a)v_{i_0} \wedge \exp(a)v_{i_1} \wedge \ldots    
 \]
 We will need slightly more general group $G_{\infty}$ of matrices $g$ with $\det g \neq 0$, which have only finite number of diagonals below the main one but any number above it. It is obvious that the above formula holds for them.
 Next we define the bosonic Fock spaces $B^{(m)}$   for $m \in \Zset$. Introduce the     formal variable $Q$ and put 
 
 \beq \label{bos}
 B^{(m)} := \Cset[Q^m; t_1, t_2, \ldots].
 \eeq
The element $Q^m\cdot 1 \in  B^{(m)}$ is called the $m$-th vacuum.

Introduce  the oscillator algebra      $\mathcal{A}$ spanned by operators $a_j, \; j \in \Zset$ and an operator $c$, called central charge subject to the identities

\[
[a_i, a_j]    =       i\de_{i, -j}\cdot c,   \;\; [c, a_j] = 0.
\]
This algebra can be represented in both  the bosonic and the fermionic Fock spaces.

 If we want to represent the oscillator algebra in the fermionic Fock space in terms of the operators $r(a)$ we see that the operator $r(I)$  (here $I$ is the identity matrix) is not well defined. So we need a regularization. We will recall its definition   (cf. \cite{KR}) for the algebra  $\bar{a}_{\infty}$.

 Let $E_{ij} \in \mathfrak{ gl}_{\infty}$ be the matrix with $1$ on the intersection of the $i$-th row and the $j$-th column.  Put $\La_j = \sum_{-\infty}^{+\infty} E_{i, i+j}$.
 
 Define   the operator $\hat{r}(E_{ij})$ by

 \[
 \hat{r}_m(E_{ij}) =   \begin{cases}
 r_m(E_{ij}), \;\; \text{if}        \;\;                     i\neq j \; \text{or}  \;\;    i=j <0\\
 r_m(E_{ii})  -id,  \;\; \text{if}   \;\;                    i\geq 0.
 \end{cases}
 \]
 and for the entire algebra  $ \bar{a}_{\infty} $ continue by linearity.
 
 Notice that the elements $\hat{r}(\La_j) \in \bar{a}_{\infty} , \; j \in \Zset$ together with the central element $c \in \bar{a}_{\infty} $ satisfy the commutation relation
 
 \[
 [\hat{r}(\La_i),  \hat{r}(\La_j)]  = i\de_{i, -j} \cdot c.
 \]
We obtain a highest weight representation of $\mathcal{A}$ in  the fermionic Fock space noticing that   the operators $\hat{r}(\La_j)$ act on the vacuum as

\[
\hat{r}_m(\La_j)  \ket{m} = 0, \; j = 1, 2, \ldots
\]
 On the other hand the elements $\hat{r}_m(\La_{-j_1}),    \ldots \hat{r}_m(\La_{-j_k})  \ket{m}$  span $F^{(m)}$ so we have a highest weight representation of the oscillator algebra.
 
 Similarly  we can define a representation of $\mathcal{A}$  in the bosonic     spaces    \eqref{bos}, using the operators 
 
 \[
 J_k = \partial_{t_k}\;\text{and}\;\; J_{-k} = k{t_{k}}\;\;   \text{for}\;\; k\geq0.
 \]
 Notice that the operators $J_k, \; k>0$ act on the vacuum $Q^m\cdot 1$ by $J_k Q^m\cdot 1 =0$ and the operators $J_{-k}, \; k>0$ acting on it span the entire bosonic space $B^{(m)}$.

 It is known  that the two representations of    $\mathcal{A}$ in the bosonic and in the fermionic Fock spaces are  equivalent, see \cite{KR}. The map between $F^{(m)}$  and $B^{(m)}$    is called
 boson-fermion correspondence Let us denote it by $\sigma_m$.

Our next goal is to  introduce the Virasoro algebra $Vir$.  It is the algebra spanned by a set of operators $L_k, \; k\in\Zset$ and $c$ (the central element) with the following commutation relation

\beq    \label{Vir}
[L_k, L_m]   =  (k-m) L_{k+m} + \frac{k^3 -k}{12} \de_{k,-m}c.
\eeq
We aim at defining two representations of this algebra.  First we  introduce the  bosonic field

\[
J(z)  = \sum_{m\in \Zset} J_mz^{-1-m}
\]
and  define  the operators $L_k$ by 

\[
\frac{1}{2}\colon  J(z)^2 \colon =  \sum_{m\in \Zset}L_m z^{-m-2},
\]
where 
\[
\colon J_k J_l \colon    =  \begin{cases} J_kJ_l,   \;  \text{if }  \; k <0\\
 J_lJ_k, \;  \text{if }   \;k >0
\end{cases}.
\]
Explicitly the operators $L_m$ are given by 

\[
L_m = \sum_{l+k = -m} lk t_lt_k  +\sum_{k=1}^{\infty} kt_k \partial_{t_{k+m}}  +
\frac{1}{2} \sum_{k+l = m} \frac{\partial^2}{\partial {t_k}   \partial {t_l}    }.
\]
 One can show that they satisfy the commutation relation of the Virasoro algebra.

 We  can define a fermionic representation of $Vir$.
 Let us take the space $\Vset$ to be spanned by     functions  $v_k(z) =z^k + \mathcal{O}(z^{k-1})  ,\;\; k\in \Zset$.   We see that the operators $d_k =-z^{k+1}\partial_z$ act on it and satisfy the commutation relation
 
 \[
 [d_k, d_m] = (k- m)d_{k+m},
 \]
 which defines centerless algebra. Then we can define action on the corresponding fermionic Fock space  $F^{(m)}$. It is easy to  show that the operators $\hat{r}(d_k), \; k \in \Zset$ and $c  = c_m$, (where $c_m$ is not important here)  define a representation of the Virasoro algebra. 
 
  One can show that the above described representations of the Virasoro algebra in  $F^{(m)}$  and $B^{(m)}$  are also equivalent, cf. \cite{KR}.

   Below we will present the formula for this equivalence in an appropriate  basis. Let us take instead of $d_k$ the elements $b_k = - \partial_z z^{k+1}$. Then

\beq \label{Bos-W}
Y_{b_k}  =  \Res_z  \Big( z^{k+1} \frac{ \colon (J(z) + \partial_z)J(z)\colon}{2}   \Big).
\eeq
 
 \bre{sub}
 
  We point out that there are subalgebras of Vir, which are isomorphic to Vir, cf. \cite{KR}. Namely let $p \in \Nset$   be fixed. Consider the elements $L_{pj}, \; j \in \Zset$.   These elements satisfy the commutation relations
 
 \[
 [L_{pk}, L_{pn}]  = p(k-n) L_{pk +pn}    + \frac{(pk)^3    -pk}{12}  c \de_{k, - n}.
 \]
 If we put     
 \[
\tilde{c}  = pc,\;      \tilde{L}_j =   \frac{1}{p} L_{pj} \; \text{for} \;    j\neq 0,   \;    \text{and}  \;  \tilde{L}_0 =      \frac{1}{p}\big( L_0 +   \frac{ p^2c - c}{24}   \big)
  \]
 we see that the new operators $\tilde{L}_j$  and $\tilde{c}$ satisfy the Virasoro commutation relations      \eqref{Vir}.  In what follows we are going to use this subalgebra.
 
 \ere

\section{Toda flows}  \label{tau}

 In this section we briefly recall the formalism of the 2D-Toda hierarchy following \cite{UT, Ta}. It is formulated in terms of difference
 operators.
 
 A   pseudo-difference operator is a finite linear combination of the form

 \[
 A = \sum_{n = -\infty}^{N} a_n(s)\La^n  \;\; \text{(operator of }  (-\infty, N]\; \text{type})
 \]
 and

 \[
 A = \sum_{n = M}^{\infty} a_n(s)\La^n  \;\;  \text{(operator of }     [M, +\infty)\;   \text{type}).
 \]
 These operators are analogs of the pseudo-differential operators for the theory of KP-hierarchy,  \cite{Ad, JM, vM, Di}. We briefly describe the construction.

Let the pseudo-difference operator  $A$ have the form 
 
 \[
 A = \La +  \sum_{n = -\infty}^{0} a_n(s)\La^n.
 \]
The operator $A$ can be conjugated to $\La$ by some operator    $W$ of the form 

\[
W = 1 + \sum_{j=0}^{\infty}   \ga_j \La^{-j},
\]
called wave operator.
 
For an operator 

\[
A  =  \sum_{-\infty}^{N} \ga_j \La^j
\]
 as usually we define $A_+$ to be the part of $A$ containing the nonnegative powers of $\La$:
 \[
 A_+ = \sum_{j=0}^{N} \ga_j \La^j
 \]
 and $A_- = A - A_+$.
 Then the Toda flows are given by 
 
 \[
 A_{t_k}   = [  (A^k)_+,     A].
 \]
 These flows act on the wave operator $W$ as
 
 \[
 W_{t_k}   =  -(A^k)_- W.
 \]

  Consider a recurrence relation of the form
 
 \[
 x w_n(x) = w_{n+1}(x)     +\sum_{j=0}^{d}\ga_j(n)w_{n-j}(n).
 \]
 It defines a   difference operator  $D$ 
 
 \[
 D = \La   +\sum_{j=0}^{d}\ga_j\La^{-j},
 \]
such that $Dw =xw$ with  $w = (\ldots, w_{n-1},   w_n, w_{n+1},  \ldots)^T$. 
 $D$	 can be conjugated to $\La$ using an operator of the form $\overline{W}  = 1 +\sum_{j=1}^{\infty }\ga_j \La^{-j}$ 	or to $\La^{-d}$ by using $W=  \sum_{j= 0}^{\infty }\de_j \La^{j}$,  $\de_0 \neq 0$. 
 
 \[
 D = \overline{W} \La \overline{W}^{-1}     =  	W\La^{-d} W^{-1}.
 \]
 One can easily show that the definition of $D^{\frac{1}{d}}$ is correct.

 The 2D-Toda hierarchy is the system of equations 
 
 \[\begin{split}
 \partial_{\bar{t}_k} D = [(D^k)_+,   D] \\
 \partial_{{t}_k} D = [(D^\frac{k}{d})_+,   D].
 \end{split}
 \]
 Notice that in our case we can assume that $\bar{t}_k =  t_{kd}$. For this reason from now on we will work only with the set of variables $t_k$. 
 


 
We will  apply this  in the situation from  \leref{recur}. It shows that    we can construct the   vector 
\[
\nu=(\ldots  \nu(s+n-1,x),  \nu(s,n,x), \nu(s+n+1,x),\ldots)^T,\; n\in \Zset.
\]
According to \eqref{recur}   we have 

\[
x\nu(s, x)  = (s+1)\nu(s+1,x) + \nu(s-d,x),
\]
which can be written as 

\[
x\nu = D\nu
\]
where the difference operator is given by

\[
D = A\La  + \La^{-d},  A = \diag(\ldots, s-1, s, s+1, \ldots).
\]


 Next we put $x= y^{d}$, which gives the formula

 \beq   \label{rec-2}
 y^d\nu(s)  =  (s+1)\nu(s+1) + \nu(s- d).
 \eeq


 There exists a tau-function $\tau (s, t, x)$ such that  the corresponding solution of the hierarchy (i.e. the coefficients of $D$) is given in terms of this function. While we don't need the explicit form of $D$ we do need the connection of $\tau(s,t,x)$ with the vector $w$. Namely  the   boson-fermion correspondence  maps a plane of Sato's Grassmannian (or wedge product)  to the tau-function.    We explain the construction more precisely. Each vacuum  $\ket{m}$ corresponds to  the vector $(w_m, w_{m-1},  \ldots)$. Then the boson-fermion correspondence maps   $\ket{m}$ to  $\tau(m, t)$.

 \section{Virasoro   constraints}

 We denote by $L_m$ the operators of the Virasoro algebra. Our approach will be a standard one -
to  obtain them from  differential operators $A_j$ acting on the functions $w(s,y)$.  It is enough to do this for $j= -1,0,1,2$    and use the identity      $[L_k, L_m]   =   (k-m)L_{k+m}$  to make induction.  Then we transport their action on the corresponding fermionic Fock space. 

Let us first define the fermionic Fock space. Put   $v_j = w(j,y)$ and

\[
\Vset =  \mathop{ \oplus }_{j \in \Zset} \Cset v_j. 
\]
Let $m $ be an integer. The fermionic Fock space $F^{(m)}$ is spanned by the wedge products 

\[
v_{i_m}\wedge v_{i_{m-1}} \wedge \ldots,
\]
where $i_{m} > i_{m-1} > \ldots$  and $i_k = k$ for $k<<m$.

 Define the operators $A_{-1},   A_0, A_1$ that act on $w(s,y)$    as follows
 
 \[
 \begin{split}
 &A_{-1} w(s,y)   =  -  \frac{1}{d}   y^{-d+1}\partial_y   w(s,y)   \\
 &   A_0  w(s,y) =       -  \frac{1}{d}    y \partial_{y}  w(s,y)   \\
  & A_1 w(s,y)  =   -  \frac{1}{d}  y^{d} [y \partial_{y}  - ds]  w(s,y).
 \end{split}
 \]
Simple computations show that these operators satisfy the following commutation relations

 \beq  \label{commA}
\begin{split}
&[A_0, A_{-1} ]   =      A_{-1}  \\
&   [A_1, A_0]   =      A_1\\
& [A_1,   A_{-1}]   = 2A_0 .
\end{split}
\eeq


  Our goal will be to present   the differential operators $A_j, \; j= -1,0,1$ as linear operators (eliminating the differentiation) in the basis $v_k$.    Then we will transport them to the fermionic Fock space $F^{(m)}$ to obtain  the Virasoro operators $L_j$.
  However we will see that with  $L_2$ the situation is more subtle   -  it does not come from any differential operator.

   First we write    $A_j, \; j=   -1,0,1$  in the space spanned by $w(n, y), \; n\in \Zset$.

 \ble{vir1}
 The following identities hold
 
 \beq  \label{A}
 \begin{split}
    (i)  \;  A_{-1}w(s,y)  &=  -    w(s-1, y)  \\
 (ii)\;\;  A_0  w(s,y)&=        - [sw(s,y)  +w(s-d-1,y)]\\
(iii)    A_1 w(s, y)
&  =  - [(s- d) w(s-d)   +w(s-2d-1) ].
 \end{split}
 \eeq
 
 \ele
  
  \proof  Using that
  
  \[
 d x\partial_x = y\partial_y.
  \]
  we obtain  
  
  \[
   A_{-1} w(s,y)= - d^{-1} y^{1-d}\partial_yw(s,y)    =   -   w(s-1,y).
  \]
 The second identity follows from the first   one by applying the recurrence relation once:
 
 \[
 \begin{split}
A_0 w(s, y) &= y^dA_{-1}w(s,y)  = -   w(s-1,y)\\
 &= -[sw(s,y)  + w(s-d-1)].
 \end{split}
 \]
 Repeat the above procedure to obtain the third relation:

 \[
 \begin{split}
  A_1 w(s, y)& = -y^d w(s-d-1)\\
 &  = - [(s- d) w(s-d)   +w(s-2d-1) ].
 \end{split}
 \]

  \qed

  \bco{oper} 
   In the basis $v_j$ the operators $A_k$ are given by

  \beq  \label{A-lin}
  \begin{split}
  (i)  \;  A_{-1}v_j  &=  -    v_{j-1}  \\
  (ii)\;\;  A_0  v_j &=        - [jv_j +v_{j-d -1}]\\
  (iii)    \; A_1  v_j &=  -(j-d) v_{j-d}-v_{j- 2d -1}. 
  \end{split}
  \eeq
    
  \eco
We want to transport the action of the   operators     $A_k$   on the vacuum

\[
\ket{m} =  v_m\wedge v_{m-1}\wedge\ldots
\]


 
 \ble{Vir0}
 We have 
 
 \beq  \label{vir0}
 \begin{split} 
 &\hat{r}_m(A_{-1}) \ket{m} =0\\
&\hat{r}_m(A_0) \ket{m}=   \frac{(m+1)m}{2}  \ket{m}\\
&\hat{r}_m (A_{1}) \ket{m} =0.\\
 \end{split}
  \eeq

 \ele
 
 \proof
 
  The first and the third identities are obvious. For the second one we need to apply the regularization procedure.

  \qed

  The operator    corresponding to Virasoro $L_2$ will be obtained following  an idea  from \cite{HH} (see also \cite{Al2}). It will be constructed as a sum of two operators, acting on the fermionic spaces $F^{(m)}$.  The first one  will be the image of some differential operator  $\tilde{A}_2$. The second one     will be $\be[\hat{r}(y^{d})]^2$, where $\be$ is some constant, which will be determined later. Notice that $\be[\hat{r}(y^{d})]^2$ is not an operator that comes from differential one.

     We define  the operator $\tilde{A}_2$  as
  
  \[
 \tilde{A}_2 =y^{2d} \big[A_0  + \al\big],
  \]
  where $\al \in \Cset$ will be determined  later.   Now put
  
  \[
L_2^F = \hat{r}_m(\tilde{A}_2)-  [\hat{r}_m(y^{d})]^2. 
  \]

  \ble{vir2}
       If $  \be =  - 1, \; \al = 2$  then
       
       \[
 L_2^F\ket{m} =0.
       \]
  \ele

  \proof

 First let us compute action of $\tilde{A}_2$ on the vectors $v_j$.  
  We use the recurrence relation.
  \[
  \begin{split}
  	&\tilde{A}_2v_j =    y^{2d}  [A_0  +\al]v_j=  \\
  	& =  -y^{2d} \big[(j   -\al)v_j  +    v_{j   -d-1} \big]. \\
  \end{split}
  \]
  Then using the recurrence relation twice we obtain
  
  \[
  \begin{split}
  \tilde{A}_2v_j &=    -y^d\big[(j   -\al)  [ (j+1)v_{j+1}  +       \mathcal{L}(j-d)\big]\\
  & =-(j-\al)(j+1)v_{j+2}  +       \mathcal{L}(j-d+1)\big],
  \end{split}
  \]  
  where the symbol $\mathcal{L}(n)$  denotes terms $v_n$ with indexes $n$ and lower.

  Then for $\hat{r}_m(\tilde{A}_2) \ket{m}$  we obtain
  
  \[
  \begin{split}
 &\hat{r}_m(\tilde{A}_2) \ket{m}\\
 & =  (m-\al)(m+1)v_{m+2}\wedge v_{m-1} \wedge \ldots \\
 & - (m-1-\al )mv_{m+1}\wedge v_{m}\wedge      v_{m-2}\wedge\ldots
  \end{split}
  \]


  For the computation of  $(\hat{r}_m(y^{d}))^2$ we will use the following formula (see, e.g. \cite{Al})

  \[
  \begin{split}
  &[\hat{r}(b)]^2 (v_m\wedge v_{m-1}  \wedge \ldots)  =  \\
  &=\sum_{-\infty}^{m}(v_m\wedge v_{m-1}  \wedge \ldots \wedge b^2 v_l \wedge\ldots)  
     + 2   \sum_{k>l} (v_m\wedge v_{m-1}  \wedge \ldots \wedge b v_k \ldots\wedge b v_l \wedge\ldots).
  \end{split}
  \]
  Using this formula we obtain   
  
  \[
  \begin{split}
  & (\hat{r}_n(y^{d}))^2  [v_m\wedge v_{m-1}\wedge \ldots]  \\
  &= (m+2)(m+1)v_{m+2}\wedge v_{m-1} \wedge v_{m-2} \ldots  + m(m+1)   v_{m } \wedge v_{m +1}\wedge v_{m-2} \ldots \\
  &+ 2(m+1) m v_{m+1} \wedge v_m \wedge v_{m-2} \ldots\\
  &  = (m+2) (m+1)v_{m+2}\wedge v_{m-1} \wedge v_{m-2} \ldots  +  m(m+1) v_{m+1 } \wedge v_{m}\wedge v_{m-2} \ldots .
  \end{split}
  \]
  
  Finally for $\hat{r}(\tilde{A}_2) + \be (\hat{r}_m(y^{d}))^2 $ we obtain 
  
  \[
  \begin{split}
  &[\hat{r}(\tilde{A}_2)   +\be (\hat{r}_m(y^{d}))^2 ]     \ket{m} =\\
  &   = \big[ (m-\al)(m+1)+ \be(m+2) (m+1)\big]v_{m+2}\wedge v_{m-1} \wedge \ldots \\
  & +  \big[(m-1-\al)m +  \be m(m+1)       \big] v_{m+1}\wedge v_{m}\wedge\ldots\\
  \end{split}
  \]
  If we want that the above terms to cancel     we have to put 
  
  \[
  \be =  - 1, \; \al = - 2.
  \]

  \qed

   \bth{bos-f}
   
 The bosonic representation of the operators $A_j, \; j =0,\pm 1$ are as follows

 \[
 \begin{split}
 Y_{A_{-1} } &=\frac{1}{d} \big[\sum_{j=d}^{\infty} jt_{j}\partial_{t_{j-d}} + \frac{1}{2}\sum_{j=1}^{d-1}jt_j (d-j)t_{d-j} \big]
 \\
 Y_{ A_0 } &=  \frac{1}{d} \sum_{j=1}^{\infty} jt_{j+d}\partial_{t_j}  \\
 Y_{A_{1} } &= \frac{1}{d}\big[ \sum_{j=1}^{\infty} jt_{j}\partial_{t_{j+d}}    -\frac{1}{2} \sum_{j=1}^{d-1} \partial_{t_j} \partial_{t_{d-j}}    \big]   
 \end{split}
 \]
 The Virasoro operators $L_j$ for $j=0,\pm 1$ are given by $L_j = Y_{A_j}$.

  \ethe

  \proof
  
  The boson-fermion correspondence    \eqref{Bos-W}       gives
  
  \[
  \begin{split}
  Y_{A_{-1}} &=  
  \frac{1}{d}\Res_y  \Big(y^{1-d} \frac{ \colon (J(y) + \partial_y)J(y)\colon}{2}  \Big)    \\
  & = \frac{1}{d}  \big[ \sum_{j=d}^{\infty} jt_{j}\partial_{t_{j-d}} + \frac{1}{2}\sum_{j=1}^{d-1}jt_j (d-j)t_{d-j}\big].
  \end{split}
  \]
  Next consider

  \[
  Y_{A_0 } =  -   \frac{1}{d}\Res_y  \Big( y \frac{ \colon (J(y) + \partial_{y})J(y)\colon}{2}  \Big)  =   -\sum_{j=1}^{\infty} jt_{j+d}\partial_{t_j}.
  \]
  
  For the computation of $Y_{A_1}$ we need the formula

  \beq \label{y^d}
  \begin{split}
  Y_{y^d } &=  -\Res_y  \Big(y^{d}J(y)    \Big)  = \partial_{t_d}.
  \end{split}
  \eeq

 Then we get

  \[
  \begin{split}
  Y_{A_1 } &=  -   \frac{1}{d}\Res_y  \Big( y^{d+1} \frac{ \colon (J(y) + \partial_{y})J(y)\colon}{2}  \Big)    -    m  \Res_y  \Big(y^{d}J(y)    \Big) \\
  & =- \sum_{j=1}^{\infty} jt_{j}\partial_{t_{j+d}}    +\frac{1}{2} \sum_{j=1}^{d-1} \partial_{t_j} \partial_{t_{d-j}}  
 +m \partial_{t_d}.
  \end{split}
  \]
    From the above formulas we obtain the explicit expressions for $Y_{A_j}, j=0, \pm 1$.
  
  The last statement follows from the commutation relations for the operators $A_j$.
  
  \qed
  
  Now we turn to $L_2^F$.
  
  \ble{L2}
  The bosonic represenation of $L_2^F$  is given by
  
  \[
  L_2 = \frac{1}{d}\big[ \sum_{j=1}^{\infty} jt_{j}\partial_{t_{j+2d}}    + \frac{1}{2} \sum_{j=1}^{2d-1} \partial_{t_j} \partial_{t_{2d-j}} + \partial_{t_d}^2\big].
  \]

  \ele
  
  \proof
  
      The same computation as above gives

  \[
  \begin{split}
  Y_{A_2 } &=  \frac{1}{d}\Res_y  \Big( y^{2d+1} \frac{ \colon (J(y) + \partial_{y})J(y)\colon}{2} + 2y^{2d} J(y)   \Big)\\
  & =\frac{1}{d}\big[ \sum_{j=1}^{\infty} jt_{j}\partial_{t_{j+2d}}    + \frac{1}{2} \sum_{j=1}^{2d-1} \partial_{t_j} \partial_{t_{2d-j}} + 2\partial_{t_{2d}}\big].
  \end{split}
  \]
From this   and from     \eqref{y^d} we find

 \[
 L_2  = \frac{1}{d}\big[ \sum_{j=1}^{\infty} jt_{j}\partial_{t_{j+2d}}    + \frac{1}{2} \sum_{j=1}^{2d-1} \partial_{t_j} \partial_{t_{2d-j}}  + \partial_{t_d}^2 \big].
 \]
 
 \qed


  We sum up the obtained results in the next theorem.

  
  
  \bth{Vir}
  The tau functions  $\tau_n$ satisfy the Virasoro constraints  
  \[
  \begin{split}
 &L_{-1}\tau_m = 0\\
 & L_0 \tau_m =  \frac{m(m+1)}{2}\tau_m \\
&L_j\tau_m     = 0, \;   j=1,2, \ldots    
  \end{split}
  \]
  \ethe

  \section{Discussion}
  
  The goal of this section is to review well  known facts about some of the above tau-functions and to put all of them in more general context.  Following  \cite{Kon, AvM1, LZ, Al2} etc. one can show that all of them   can be represented as matrix integrals of the form

  \beq \label{MI}
  Z(M)  = \det(M)  C\int \exp(-\Tr(\frac{\Phi^{d+1}}{d+1}       +M \Phi +(m+1)\log(\Phi)))  [\dd \Phi]. 
  \eeq

 Here  $M$ is a diagonal matrix $M = \diag( \mu_1, \mu_2, \ldots, \mu_N)$,   $\det M \neq 0$, $C\neq 0$ is some constant, which is irrelevant for the tau-function.   The integration is taken on the space of  Hermitian $N\times N$ matrices. The connection with the tau function $\tau_m(t_1, t_2, \ldots)$   is the following.    Let us define the parameters $t_1, t_2, \ldots$ by 
  the well known  Miwa parametrization:
 
 \[
 t_j = \frac{1}{k}\Tr M^{-j}.   
 \] 
Then  it is known that the function $\tau(t) = Z(M)$ up to irrelevant multiplicative constant. The parametrization for any fixed number of variables $t_1, t_2, \ldots$  does not depend on the size $N$ of the matrices provided it is large enough. For more details see \cite{LZ}.

To be precise, it is proved  for few cases. However they all can be done more or less following the same pattern.

Some of these matrix integrals describe beautiful algebraic geometry - intersection numbers on different moduli spaces. Namely introduce the function (it is rather asymptotic expansion)

\[
F = \log(Z(t)) =  \sum_{\al} B_{\al} t^{\al}.
\]

Then  the coefficients $B_{\al}$  give intersection numbers in different situations.

 The case of $d=2$ is the most important. When $m= -1$ it corresponds to the Kontsevich integral \cite{Kon, Wi}, which describes the intersection theory on all moduli spaces   $\mathcal{M}_{g,n}$   of compact Riemann surfaces of genus $g$ and $n$ marked points. The case of $m= 0$  describes the intersection theory of open Riemann surfaces. These are Riemann surfaces from which a finite number of discs have been deleted. For more details  of this very recent theory see \cite{Al2, Bu} and the references therein.
 
 Other cases of interest include arbitrary $d$ and $m=-1$. These are the    partition function for $r$-spin structures of type $A$ (here $r=d+1$).  These were introduced by Witten \cite{Wi} who also formulated the conjecture   that the corresponding intersection theory is governed by the model \eqref{MI}   with $d>2$  and $m=-1$. The conjecture was proved in \cite{FSZ}.
 
 Recently there was a suggestion \cite{BD}   that the case of $m=0$ will produce the open analog for the $r$-spin structures.  
 
 It is natural to ask if the rest of the tau-functions have some algebro-geometric meaning. This is not clear even for the case $d=2$.  
 
 Finally let us recall that instead of integer $m$ we can take any complex number (see   \eqref{eps}). Then we will obtain the extended   Toda hierarchy, see \cite{Car}.  It contains other flows.   
They  deserve further studies.
 

	\pdfbookmark[1]{References}{ref}
	

\end{document}